\documentclass[conference,10pt]{IEEEtran}

\usepackage{amsmath,amsfonts}
\usepackage{algorithmic}
\usepackage{algorithm}
\usepackage{array}
\usepackage[caption=false,font=normalsize,labelfont=sf,textfont=sf]{subfig}
\usepackage{textcomp}
\usepackage{stfloats}
\usepackage{url}
\usepackage{verbatim}
\usepackage{graphicx}
\usepackage{cite}

\usepackage{tikz}
\usetikzlibrary{shapes.geometric, arrows.meta, positioning}

\tikzstyle{block} = [rectangle, rounded corners, text centered, draw=black, fill=gray!10, minimum width=6cm, minimum height=1cm]
\tikzstyle{decision} = [diamond, aspect=2, text centered, draw=black, fill=gray!10]
\tikzstyle{arrow} = [thick, -{Latex[length=3mm]}]

\begin{document}

\title{Swarm Network-as-a-Service (SNaaS)}

\author{
\IEEEauthorblockN{Balsam Alkouz, Osama Amin, and Basem Shihada}
\IEEEauthorblockA{
Computer, Electrical and Mathematical Sciences and Engineering (CEMSE) Division\\
King Abdullah University of Science and Technology (KAUST), Saudi Arabia\\
Email: \{balsam.alkouz, osama.amin, basem.shihada\}@kaust.edu.sa}
}



\maketitle

\begin{abstract}
Emerging on-demand connectivity scenarios increasingly require networking solutions with stringent service-level guarantees. We propose Swarm Network-as-a-Service (SNaaS), a service-oriented framework that leverages fleets of drones to provide on-demand connectivity at scale. SNaaS explicitly models drone-to-device and drone-to-drone interactions as composable services, enabling consumers to request connectivity through Service-Level Agreements (SLAs). We formalize atomic and composite SNaaS services, present an SDN-inspired architecture that integrates the service-oriented triad of provider, consumer, and registry. We introduce a composition framework that orchestrates drones into end-to-end services. Within this framework, we define and analyze three composition strategies, i.e., direct, clustered, and parallel, and propose a queuing-theory-based heuristic for selecting the most suitable strategy under varying load conditions. A dedicated enforcement module continuously monitors queue stability and SLA latency, adaptively reconfiguring the swarm when violations occur. Experiments using real air-to-ground measurements show that the framework consistently outperforms fixed compositions, achieving lower latency, fewer SLA violations, and smoother adaptation as load and swarm size increase. 
\end{abstract}


\section{Introduction}


The rise of smart and cognitive cities, coupled with the rollout of 6G networks enabling ultra-low latency and massive connectivity, is driving unprecedented data growth and demanding networking infrastructures that can meet extreme service-level requirements at scale \cite{galal2023cognitive,dang2020should, zhang2022big}. Importantly, this need for on-demand, high-capacity networking is not limited to futuristic cities. It is already evident in scenarios such as \textit{temporary high-density events}. For example, the FIFA World Cup create connectivity demand that spikes far beyond the capacity of fixed infrastructure. Current industry practice relies on Cells on Wheels (COWs), which are costly, rigid, and limited in scalability \cite{ghoshal2023performance}. Similarly, in \textit{infrastructure-less environments} such as regional construction sites, offshore oil rigs, or disaster zones, connectivity is typically provided by satellites. These are slow to deploy, expensive, and unsuitable for low-latency, high-throughput requirements \cite{9210567}.

Drones or Unmanned Aerial Vehicles (UAVs) provide a compelling alternative. They are agile, quickly deployable, and capable of forming airborne networks on-demand. Yet, due to bandwidth, hardware, and scheduling constraints, a single UAV can typically support only a limited number of concurrent devices (on the order of a few tens), making it insufficient for high-demand scenarios \cite{xu2021throughput}. To provide meaningful capacity, \textit{drones must operate as a swarm}, collectively serving thousands of devices. Furthermore, in infrastructure-less environments, drones can act as relays, extending connectivity across vast or obstructed terrains. This introduces a new paradigm: swarm-based, on-demand networking infrastructure.

Research in Non-Terrestrial Networks (NTNs) has explored the role of drones, balloons, and satellites in providing connectivity \cite{ammar2024depth, zhou2023outage}. However, existing efforts have largely focused on algorithmic or architectural optimizations without framing the problem in terms of services \cite{javaid2023communication}. We argue that the \textit{service paradigm }provides a natural abstraction by separating the \textit{functional property}, which defines the provision of connectivity services within a target area, from \textit{non-functional properties}, which describe the Quality of Service (QoS) requirements. In this context, key QoS requirements such as end-to-end latency, aggregate throughput, and signal reliability can be explicitly specified in Service-Level Agreements (SLAs). Modeling \textbf{Swarm Network-as-a-Service (SNaaS)} allows consumers to request connectivity based on these guarantees, without concerning themselves with the underlying deployment, coordination, or management of the UAVs. 


Viewing SNaaS through the service paradigm introduces challenges in \textit{allocation}, \textit{composition}, \textit{scaling}, and \textit{robustness}, all of which require rethinking traditional networking under a service-oriented lens. In particular, \textit{composition} concerns assembling drone-to-device and drone-to-drone links into end-to-end services that satisfy global QoS guarantees such as latency, throughput, and resilience. \textit{This work focuses on the composition problem, optimizing how individual assignments are orchestrated into reliable, QoS-aware connectivity services}.\looseness=-1

Prior work on Swarm-based Drone-as-a-Service (SDaaS) modeled swarms for delivery tasks, optimizing routes in skyway networks under payload and charging constraints 
\cite{alkouz2022density}. SDaaS assumes fixed paths and focuses on delivery deadlines \cite{alkouz2023failure}. In contrast, SNaaS addresses \textit{connectivity}, operating in \textit{open, dynamic airspace} where drones must compose communication links to devices and each other. The optimization goals also shift from minimizing delivery time to ensuring QoS guarantees such as throughput, latency, and resilience. \looseness=-1

This paper lays the groundwork for SNaaS by formally defining the paradigm, presenting a service-oriented architecture, and addressing the fundamental challenge of composition in a confined area. The goal is to establish SNaaS as a foundation for a new line of research at the intersection of service computing and non-terrestrial networking.

The contributions of this work are as follows:


\begin{itemize}
    \item We introduce SNaaS for UAV-assisted connectivity, formally defining atomic and composite services that characterize its functional and non-functional attributes.
    \item We design an integrated Service-Oriented Architecture and Software-Defined Networking architecture that enables discovery, composition, and enforcement of swarm connectivity services while abstracting away swarm coordination from consumers.
    \item We propose and evaluate a queueing theory driven composition and enforcement framework that dynamically selects and adapts service compositions to preserve latency and stability guarantees under varying load.
    \item We validate the framework using real air-to-ground measurements, demonstrating efficient performance and smooth adaptation across varying load and swarm scales.
\end{itemize}

\subsection{Motivating Scenario}
\begin{figure}
    \centering
    \includegraphics[width=0.8\linewidth]{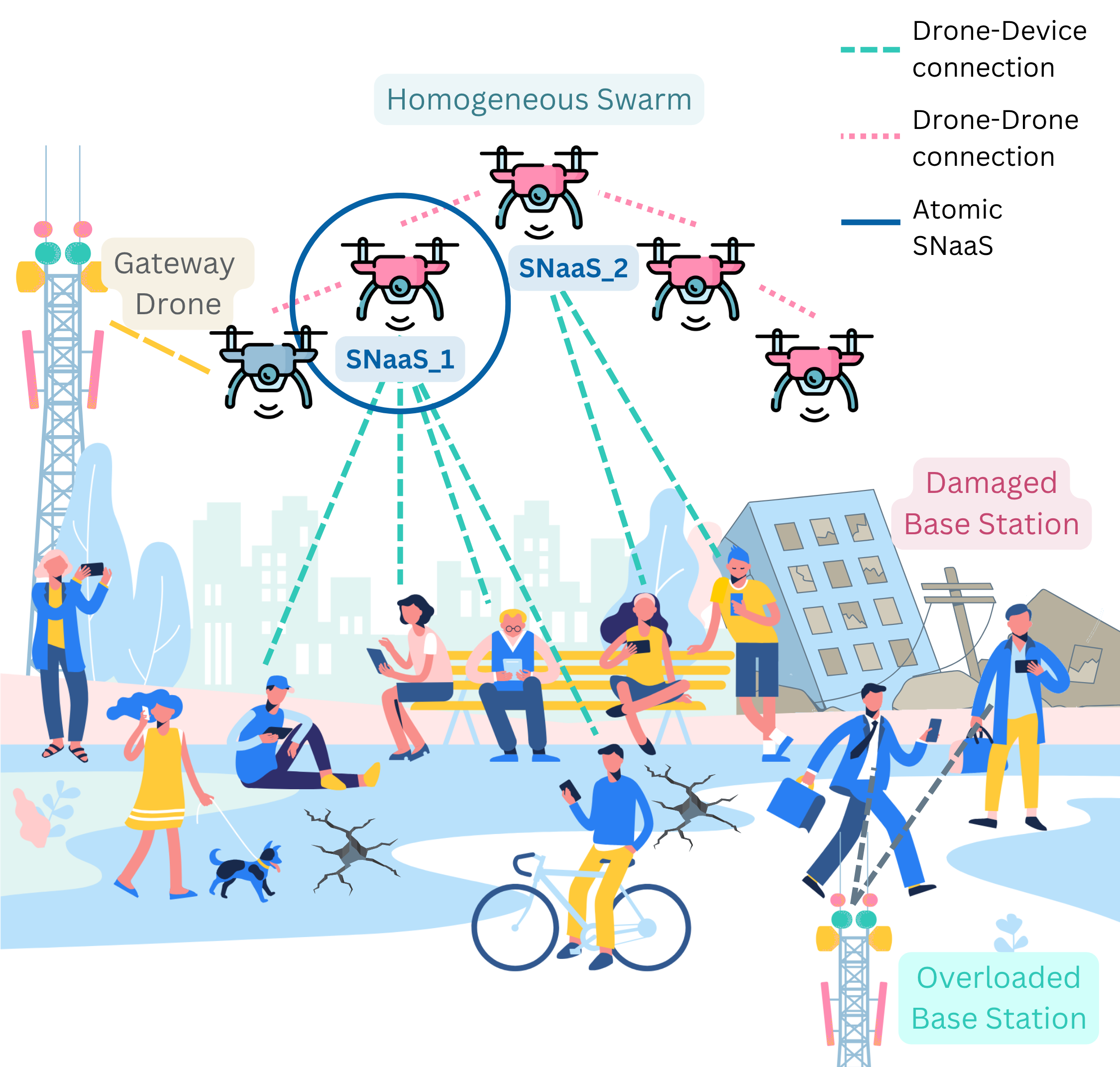}
    \caption{Illustration of an SNaaS deployment in a disaster recovery scenario. Devices connect to nearby drones (drone-to-device links), drones forward data among themselves (drone-to-drone links), and traffic exits through a gateway drone or ground station to reach the core network.}
    \label{fig:snaas_scenario}
\end{figure}

Consider a disaster recovery operation where terrestrial communication infrastructure has been destroyed. First responders and civilians in the affected area require immediate, reliable connectivity to coordinate rescue activities, transmit critical data, and maintain communication with external agencies. Terrestrial solutions are too slow to deploy, and satellites are costly and unable to provide the required low-latency, high-throughput connections. To bridge this gap, a swarm of drones is deployed to act as an airborne communication infrastructure.

In this setting, three layers of connectivity emerge, as shown in Fig.~\ref{fig:snaas_scenario}. At the first layer, \textit{devices} (e.g., mobile phones, sensors, laptops) connect to nearby drones through \textit{drone-to-device links}. At the second layer, drones communicate with each other via \textit{drone-to-drone links}, where drones forward traffic among themselves forming a multi-hop aerial relay network. At the third layer, one or more \textit{gateway drones} act as the \textit{bridge to the core network}, forwarding traffic from the swarm to terrestrial backhaul infrastructure such as fiber, 5G base stations, or satellite uplinks.

For tractability, we assume a fleet of homogeneous drones with identical communication range, energy capacity, and throughput. The drones are positioned at fixed hovering locations in the affected area, ensuring predictable coverage. Each drone can serve only a limited number of devices at once, and every device must be assigned to exactly one drone. 
We also assume that drones approaching battery depletion are proactively replaced by standby drones that take over their hovering positions, with a brief overlap period during which both drones operate concurrently to prevent service disruption. Energy optimization is outside the scope of this paper and will be considered in future work.

We define an \textit{atomic SNaaS service} as the bundle of one drone serving its allocated devices together with the drone-to-drone relay links necessary to forward data toward the gateway. To achieve end-to-end connectivity across the deployment, \textit{multiple atomic services must be composed into a single, cohesive service}. The central challenge therefore lies in designing efficient \textit{composition strategies} that \textit{orchestrate} these atomic services into an end-to-end network. Unlike allocation, which assigns devices to individual drones, composition determines how the resulting assignments are connected and coordinated to \textit{meet global QoS requirements}. In this paper, we focus specifically on optimizing service composition to minimize end-to-end \textit{latency}, subject to constraints on drone capacity and coverage. 

The difficulty of composition in SNaaS arises from the trade-offs between different ways of orchestrating atomic services. A \textit{direct composition}, where each drone forwards traffic straight to the gateway, minimizes hop count but quickly creates congestion at the gateway drone. 
\textit{Clustered compositions} reduce congestion by aggregating traffic at cluster-head drones, but require careful coordination to prevent bottlenecks at the heads. \textit{Parallel compositions} distribute traffic across multiple relay paths, but each additional hop increases end-to-end latency, making long relay chains costly despite their load-balancing benefits. These conflicting trade-offs make it difficult to determine which composition style is best suited for a given deployment. 

\section{Related Work}

UAVs have emerged as core enablers in NTNs, acting as airborne base stations, relays, or edge nodes to extend coverage and capacity in 5G/6G environments~\cite{mozaffari2019tutorial}. Prior studies focus on optimizing placement, routing, and spectrum reuse to improve throughput, latency, and energy efficiency~\cite{mozaffari2018beyond,sun2017latency,su2023energy,zhang2021energy}. For example, joint drone–user association and 3D positioning have been studied using optimal, greedy, and distributed learning-based schemes \cite{el2021optimal, el2019learn}. Some recent efforts integrate UAVs into 6G architectures to enable dynamic control and network slicing~\cite{wei2024hierarchical, ebrahimi2024resource,ammar2025maritime}. However, these works treat UAVs as network components rather than \emph{services}. 


The concept of Drones-as-a-Service (DaaS) has been applied mainly to delivery and mission execution~\cite{hamdi2025drone}. Existing frameworks in this domain define UAV services for logistics, package routing, and task assignment, often using cloud-based orchestration \cite{alkouz2021service}. Service composition and optimization in these delivery frameworks typically aim to minimize distance, energy consumption, and delivery delay \cite{lee2021package, lee2024reactive}. 

Swarm-based UAV research demonstrates how coordinated fleets outperform individual drones in coverage, fault tolerance, and scalability~\cite{alkouz2024signal, javed2024state, cao2024computational}. Cooperative control, formation flight, and multi-agent decision making have been extensively studied using consensus and reinforcement learning frameworks~\cite{alkouz2020formation, venturini2021distributed}. Our approach leverages these coordination insights but applies them to a \emph{service composition context}: drones are orchestrated into composable, SLA-compliant connectivity chains rather than purely cooperative formations.

To situate our work within this line of research, it is important to distinguish between delivery swarm services and connectivity-oriented swarm networking services. In  SDaaS, the service is explicitly defined as \emph{the delivery of packages over a skyway segment}, and composition occurs when a package cannot be delivered via a single segment \cite{alkouz2020swarm}. SDaaS therefore composes \emph{segments} in a skyway network \cite{alkouz2022density}. Crucially, the swarm membership is fixed because the number of drones is determined by the package set in the request. In contrast, SNaaS composes \emph{swarm members themselves}. Here, the service is connectivity, and when a single drone cannot satisfy QoS (e.g., latency requirements) additional drones are incorporated as infrastructure elements rather than as carriers. This makes the two paradigms fundamentally distinct in both service definition and composition logic.

Several studies label their architectures as ``UAV-as-a-Service'' or ``Networking-as-a-Service'' frameworks. For example, D3S~\cite{nait2021towards} introduces a four-phase pipeline (Demand–Decision–Deployment–Service) for adaptive aerial coverage, while other efforts propose service-based UAV architectures~\cite{bekkouche2020service} or integrate SDN with queueing models to improve control plane efficiency~\cite{abir2023software}. Although these works describe themselves as \emph{service-centric}, they \textit{operate primarily at the infrastructure or control layer}: UAVs are optimized as network components, not provisioned as consumer-facing services with measurable guarantees. In contrast, our SNaaS framework adopts a service-computing perspective~\cite{bouguettaya2017service, hamdi2025drone}, exposing swarm connectivity as an SLA-governed service.

Queueing theory has been widely used to analyze delay and stability in wireless and UAV networks~\cite{saad2009selfish}. Models such as M/M/1 and M/M/$m$ have characterized SDN controller delays, link utilization, and UAV scheduling~\cite{abir2023software}. Our work extends this foundation by embedding queueing models into the \emph{service composition process} itself. Rather than evaluating network latency post hoc, SNaaS uses queueing parameters to guide composition selection, enforce stability, and satisfy consumer SLAs dynamically.


\section{Swarm Network-as-a-Service Model}
We now formalize the SNaaS paradigm. Let $D = \{d_1, d_2, \dots, d_m\}$ denote the set of entry drones and $U = \{u_1, u_2, \dots, u_n\}$ the set of devices in the operating environment. Each device is assumed to be connected to exactly one drone, following a fixed allocation policy (e.g., assignment to the nearest drone within range). A set of gateway drones $G = \{g_1, g_2, \dots, g_\ell\}$ provides the bridge between the swarm and the terrestrial backhaul, with each gateway capable of forwarding traffic out of the aerial network.

We distinguish three functional roles for drones in the swarm. 
\emph{Entry drones} are directly connected to devices, serving as the access points for traffic entering the swarm. 
\emph{Relay drones} forward traffic toward gateways; they may either act purely as forwarders or simultaneously serve as entry drones for their own allocated devices. 
\emph{Gateway drones} act as the bridge between the swarm and the terrestrial backhaul (e.g., fiber, satellite, or 5G base station).

\subsection{Atomic SNaaS Service}
An \textit{atomic SNaaS service} represents the smallest unit of service delivery in the system. It consists of a single drone $d_j \in D$, the set of devices allocated to it, and the relay links required for $d_j$ to forward its traffic toward the gateway $g$. Formally, we denote:
\[
SNaaS_{j} =
\underbrace{\{(u_i, d_j) \mid u_i \in U_j\}}_{\text{device-to-drone links}}
\;\cup\;
\underbrace{\{(d_j, d_k) \mid d_k \in N_j\}}_{\text{drone-to-drone links}}.
\]
where $U_j \subseteq U$ is the set of devices allocated to $d_j$, and $N_j$ is the set of neighbor drones to which $d_j$ can forward data.

\subsection{Composite SNaaS Service}

While single SNaaS instances capture local connectivity, end-to-end networking requires composing multiple instances into a cohesive service. A \textit{composite SNaaS service} is therefore defined as the union of all SNaaS instances across the deployed swarm $D$, such that all devices are connected to the gateway:

\[
SNaaS_{D} = \bigcup_{d_j \in D} SNaaS_{j},
\]
A composite service is \textit{valid} if, for every device $u_i \in U$, there exists a path $(u_i, d_j, \dots, g)$ consisting of device-to-drone and drone-to-drone links that terminates at the gateway.



\subsection{QoS Objective}

The quality of a composite SNaaS service is measured by global QoS metrics such as end-to-end latency, coverage, and throughput. In this work, we focus on minimizing latency while ensuring that capacity constraints are respected:
\[
C_d = \frac{\mu}{\lambda},
\]
where \(\mu\) is the drone’s service rate and \(\lambda\) is the per-device arrival rate.

\section{SNaaS Architecture}

\begin{figure}
    \centering
    \includegraphics[width=0.9\linewidth]{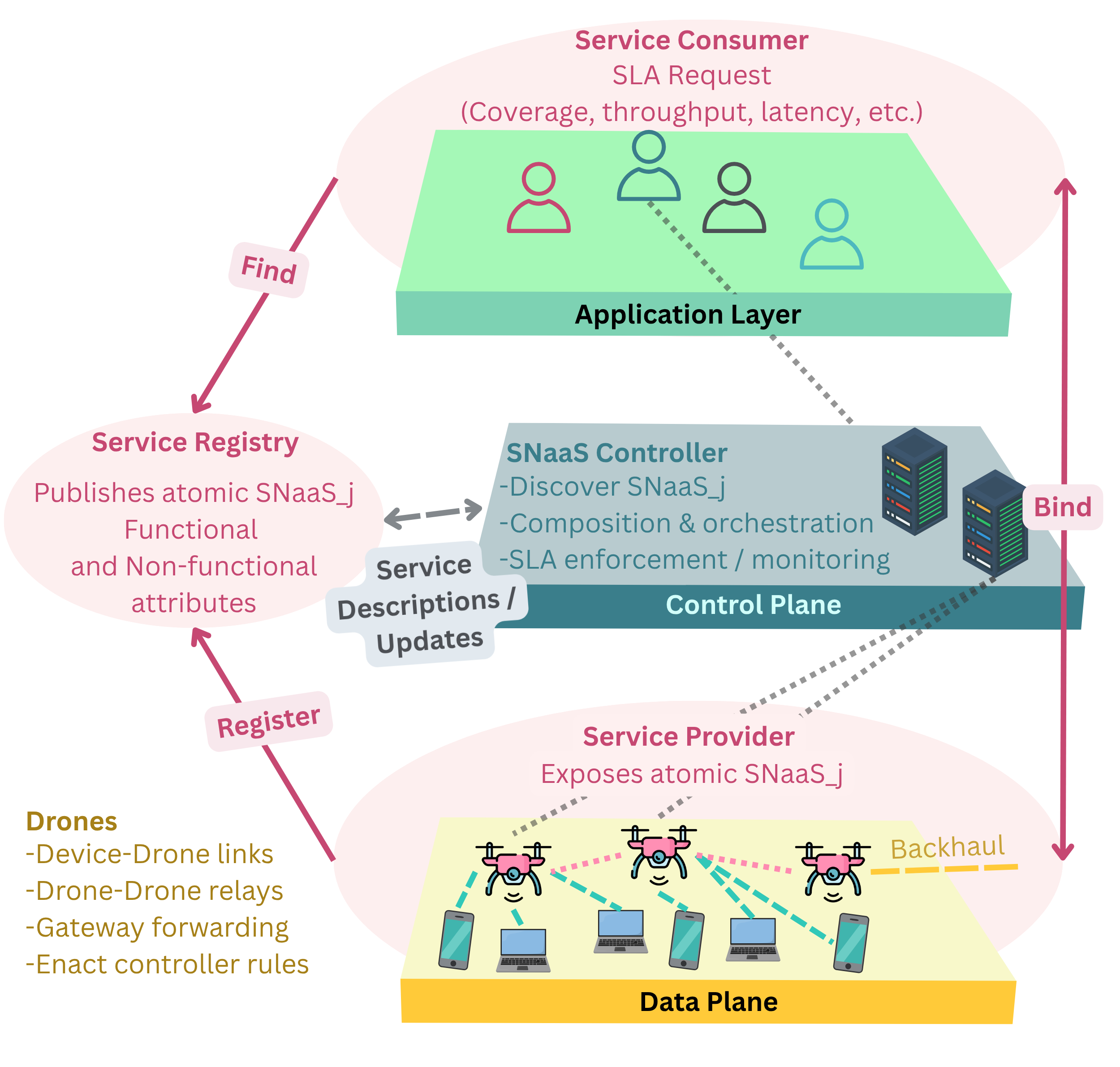}
    \caption{SNaaS architecture integrating the Service-Oriented Architecture (SOA) triad with the Software-Defined Networking (SDN) paradigm.}
    \label{fig:architecture}
\end{figure}

We model the architecture of SNaaS by integrating the \textit{Service-Oriented Architecture (SOA) triad} (provider, registry, consumer) \cite{mackenzie2006reference} with the \textit{Software-Defined Networking (SDN) paradigm} (application layer, control plane, data plane) \cite{kreutz2014software}. Integrating SOA with SDN enables SNaaS to be both consumable and enforceable: SOA exposes connectivity as a service through SLA-based abstractions, while SDN provides the control mechanisms to compose and enforce these services over the swarm. Together, they bridge the gap between high-level service requests and low-level drone orchestration.

\subsection{SOA Triad in SNaaS}
In the SNaaS context, the \textit{service provider} corresponds to the swarm operator. Each drone exposes an atomic SNaaS instance, which consists of the devices allocated to it and the relay links it maintains with its neighbors. These atomic services are published to the \textit{service registry}, which functions as the directory of available connectivity services. The registry holds both functional properties, such as neighbor relations and gateway reachability, as well as non-functional attributes, such as latency, throughput, and coverage. The service consumer interacts only with the registry, submitting connectivity requests that include desired QoS parameters or SLA requirements. 

\subsection{SDN Layer Mapping}



The SOA roles map directly onto the SDN layers: at the \textit{application layer}, consumers submit SLA-driven connectivity requests while remaining agnostic to swarm details; the \textit{control plane} functions as the SNaaS management system, discovering atomic services, composing them into end-to-end services, and enforcing QoS and capacity constraints through continuous monitoring and adaptation. While physically distributed across drones, the control plane is logically centralized, in the same spirit as SDN controllers. The \textit{data plane} is realized by the drones themselves, which execute the control decisions by establishing device-to-drone and drone-to-drone links and forwarding traffic to deliver the requested connectivity.

\subsection{Interaction Flow}

The interaction follows a simple flow: a consumer submits an SLA request, the control plane retrieves atomic SNaaS descriptions from the registry, computes a composition, and installs the resulting configuration across the drones. The registry serves as a passive catalog, while the control plane actively composes, enforces, and updates services based on swarm state. The data plane executes these decisions, and monitoring feedback enables continuous SLA validation and recomposition when needed. This architecture departs from prior cloud–edge–drone models \cite{hamdi2025drone} by embedding SDN-style plane separation within an SOA framework, enabling a clear distinction between service abstraction, orchestration, and execution. As a result, SNaaS is well suited to infrastructure-less scenarios where swarms must self-organize to deliver connectivity. \looseness=-1

\section{Queueing-Based SNaaS Composition Framework}

\begin{figure*}
    \centering
    \includegraphics[width=\linewidth]{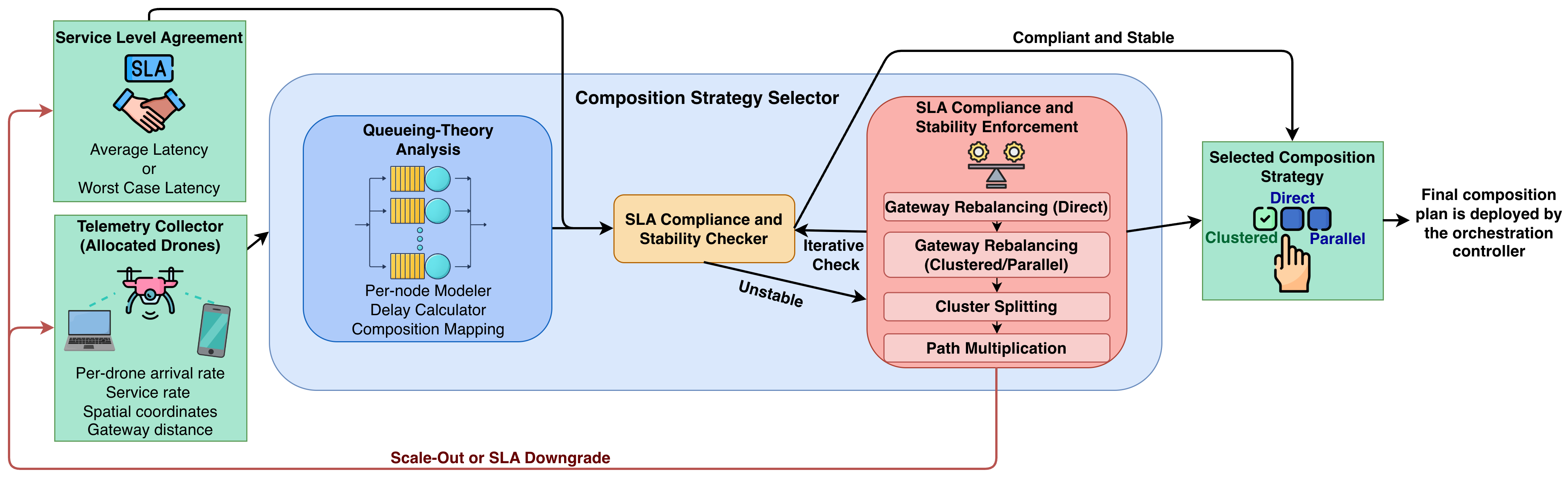}
    \caption{Queueing-Based SNaaS Composition Framework}
    \label{fig:framework}
\end{figure*}


The composition framework defines how atomic SNaaS services are orchestrated into end-to-end connectivity services (Fig. \ref{fig:framework}). It consists of three core modules: (i) \textbf{Composition strategies} (direct, clustered, and parallel) that define different orchestration patterns; (ii) \textbf{A queueing-theory-based heuristic} that evaluates each strategy under current load and topology conditions to select the one that best satisfies latency and stability requirements; and (iii) \textbf{An SLA compliance and stability enforcement module} that ensures contractual guarantees are maintained by adaptively reconfiguring the swarm when queues become unstable or latency targets are violated. \looseness=-1

The input to the framework is the current swarm state, including the set of drones, their capacities, device-to-drone allocations, and gateway availability, together with the SLA specified by the consumer. The expected output is a composed SNaaS service: a plan that defines the drone-to-drone forwarding paths from entry drones to gateways and the scheduling policy for traffic forwarding, such that the consumer’s SLA constraints are satisfied.\looseness=-1



\subsection{SNaaS Composition Strategies}

\begin{figure}[h]
    \centering
    \includegraphics[trim={0 4cm 0 3.5cm},clip, width=0.7\linewidth]{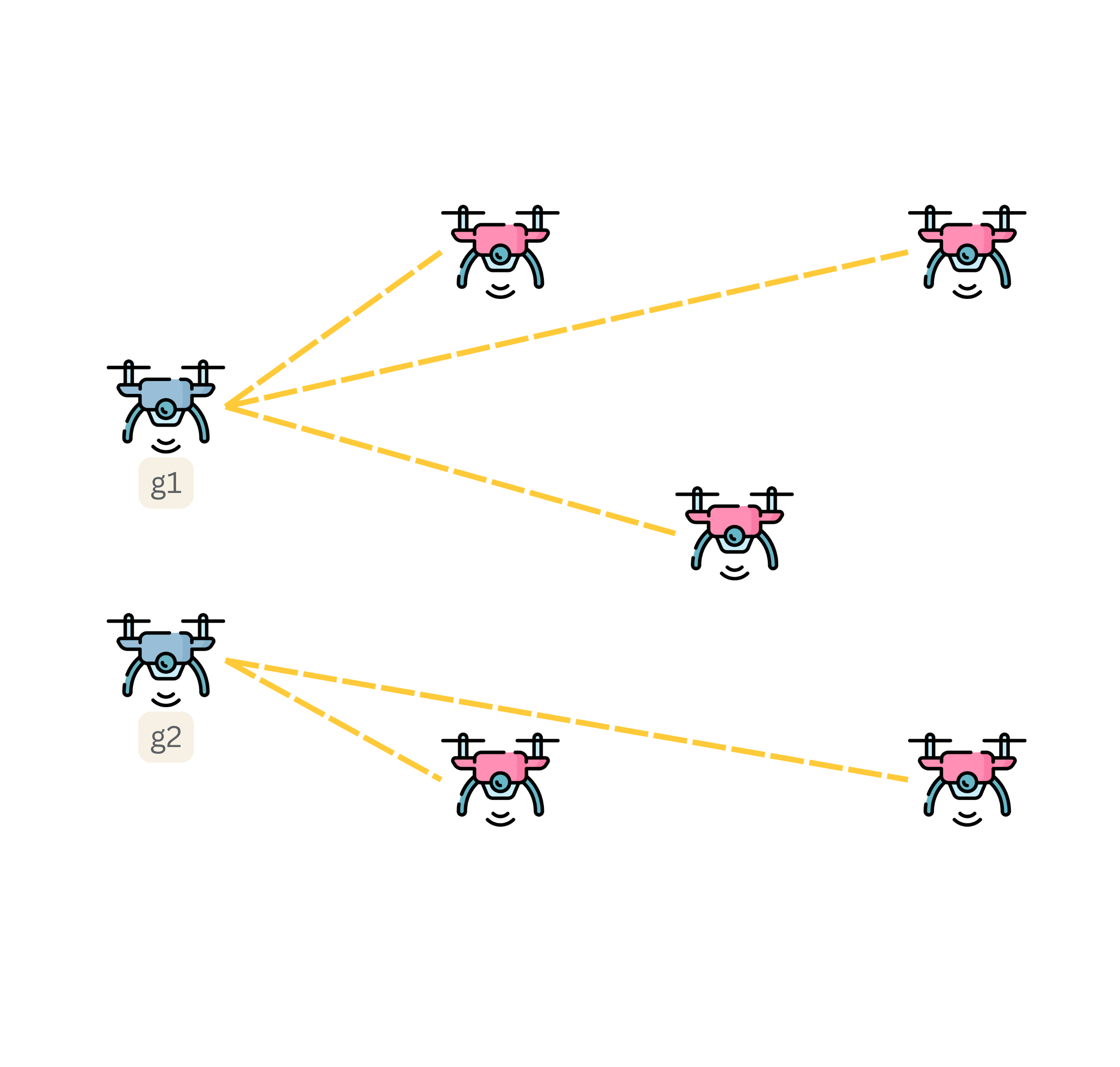}
    \caption{Direct composition: each entry drone forwards traffic directly to the gateway.}
    \label{fig:direct}
\end{figure}

\begin{figure}[h]
    \centering
    \includegraphics[trim={0 3cm 0 2cm},clip, width=0.7\linewidth]{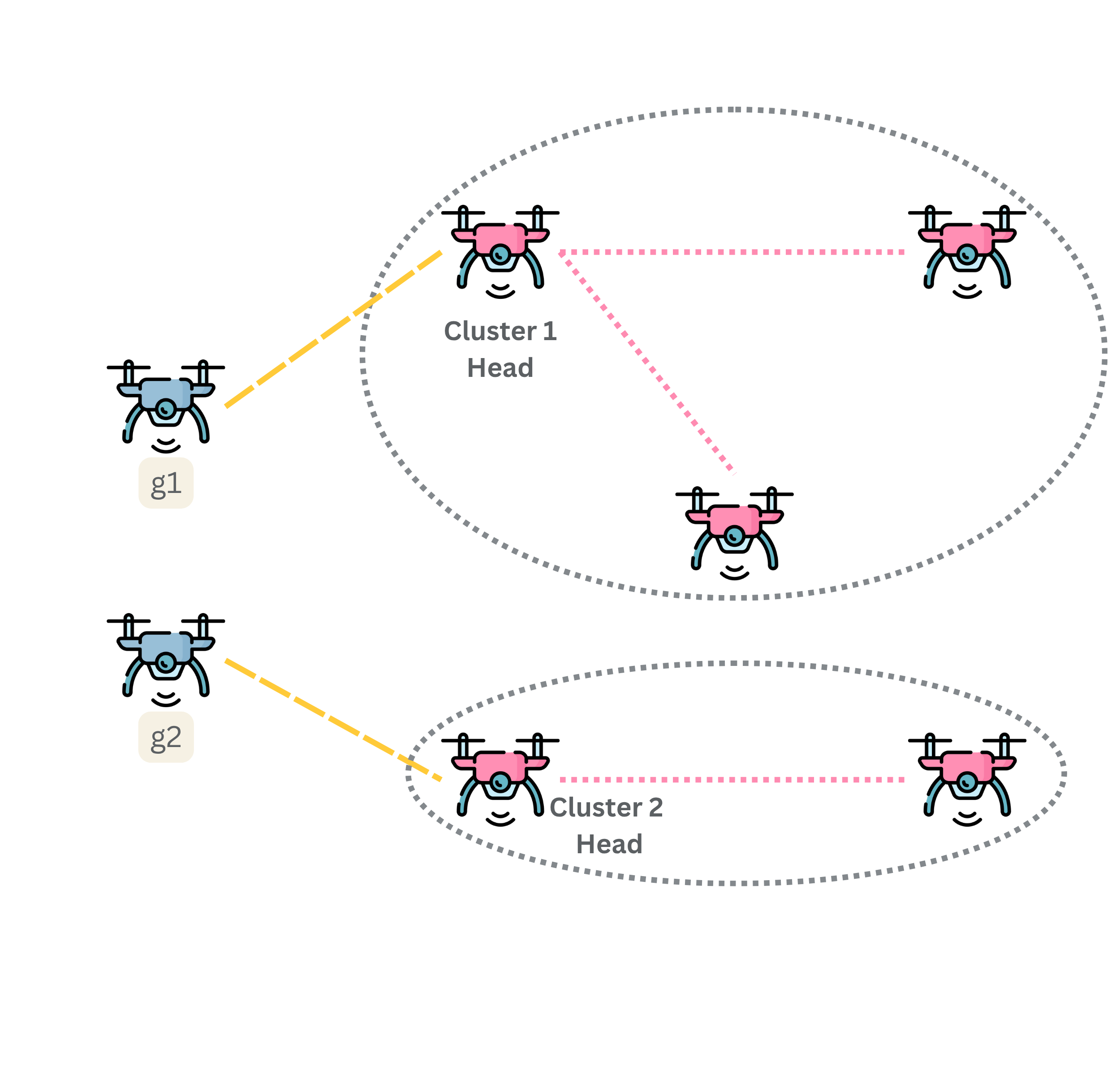}
    \caption{Clustered composition: traffic is aggregated at cluster heads before reaching the gateway.}
    \label{fig:clustered}
\end{figure}

\begin{figure}[h]
    \centering
    \includegraphics[trim={0 4cm 0 3.5cm},clip, width=0.7\linewidth]{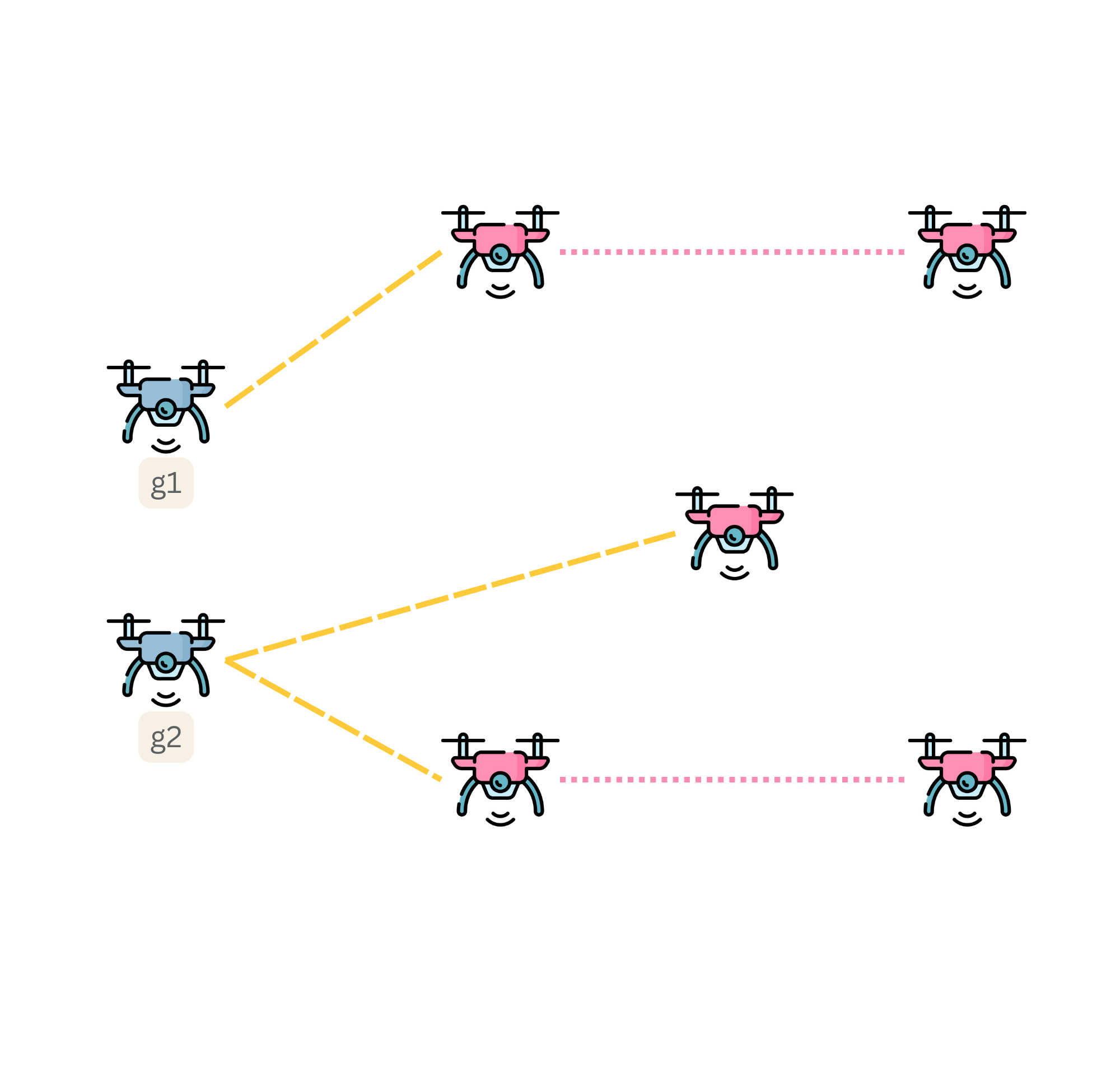}
    \caption{Parallel composition: traffic is distributed across multiple relay paths.}
    \label{fig:parallel}
\end{figure}

We introduce the composition strategies that define how drones can be orchestrated to deliver end-to-end connectivity under different load and topology conditions. We identify three fundamental ways of composing atomic SNaaS services. Each type is motivated by a distinct deployment scenario, realized by a concrete algorithm for constructing the composition, and characterized by specific performance tradeoffs. Figures \ref{fig:direct}–\ref{fig:parallel} illustrate the resulting service graphs. 

\subsubsection{Direct Composition}
In direct composition, all entry drones forward traffic directly to the gateway without intermediate relays. This configuration minimizes hop count and provides the lowest latency under light traffic. However, because every entry drone injects traffic into the gateway, congestion quickly builds up under moderate to heavy loads.

Algorithm~\ref{alg:direct} describes how entry drones are assigned to gateways in direct composition. The key decision occurs in the assignment step [Line 3], where each entry drone $d_j$ selects a gateway $g_k$ based on a weighted rule that combines proximity and current gateway load. Setting $\alpha=1$ reduces the rule to nearest-gateway selection, while $\alpha=0$ corresponds to pure load balancing. Different values of $\alpha$ enable hybrid policies that balance both factors. 

\begin{algorithm}[h]
\caption{Direct Composition Assignment}
\label{alg:direct}
\begin{algorithmic}[1]
\STATE Identify the set of gateway drones $G = \{g_1, g_2, \ldots, g_\ell\}$.
\FOR{each entry drone $d_j$}
    \STATE Assign a gateway $g_k$ by solving:
    \[
    g_k = \arg\min_{g \in G} \; \big( \alpha \cdot \text{dist}(d_j, g) + (1-\alpha) \cdot \text{load}(g) \big).
    \]
    \STATE Forward all device traffic assigned to $d_j$ along $(d_j,g_k)$ in FIFO order.
\ENDFOR
\end{algorithmic}
\end{algorithm}

\textbf{Tradeoffs.} Direct composition is efficient for small-scale or lightly loaded deployments, as it minimizes delay by avoiding relays. Its main limitation is scalability: the gateway quickly becomes a bottleneck as traffic increases. When multiple gateways exist, performance depends heavily on how effectively traffic is distributed among them.

\subsubsection{Clustered Composition}

In clustered composition, multiple drones act as \textit{cluster heads} that aggregate traffic from surrounding drones before forwarding to a gateway. This reduces congestion at the gateway by distributing load across intermediate heads. The main challenge is selecting suitable cluster heads that balance proximity and traffic load, and choosing an appropriate gateway per cluster.

Algorithm~\ref{alg:clustered} first fixes the number of clusters from capacity [Line~2] and forms spatial clusters via $k$-means [Line~3]. For each cluster, the \emph{gateway selection} step [Line~5] chooses $g^*$ using a weighted rule that balances the centroid’s proximity to each gateway and the gateways’ current load. Given $g^*$, the \emph{head selection} step [Line~6] picks $d^*$ that is both well-placed relative to $g^*$ and not overloaded. Finally, members attach to $d^*$, and $d^*$ connects directly to $g^*$ [Line~7].

\begin{algorithm}[h]
\caption{Clustered Composition Assignment}
\label{alg:clustered}
\begin{algorithmic}[1]
\STATE Identify the set of gateway drones $G = \{g_1, g_2, \ldots, g_\ell\}$.
\STATE Determine the number of clusters \[
k = \max\!\left( |G|,\; \left\lceil \frac{|U| \lambda}{\mu} \right\rceil \right),
\]
where $|U|$ is the total number of devices, $\lambda$ is the per-device arrival rate, $\mu$ is the per-drone service rate.

\STATE Apply $k$-means to the spatial positions of non-gateway drones, yielding 
clusters $\{C_1, C_2, \ldots, C_k\}$ with centroids $\{c_1,\ldots,c_k\}$.
\FOR{each cluster $C$ with centroid $c$}
    \STATE \textbf{Gateway selection:} choose the gateway for this cluster as
    \[
    g^* = \arg\min_{g \in G} \; \big( \alpha \cdot \text{dist}(c, g) + (1-\alpha) \cdot \text{load}(g) \big).
    \]
    \STATE \textbf{Head selection:} choose the cluster head as
    \[
    d^* = \arg\min_{d \in C} \; \big( \alpha \cdot \text{dist}(d, g^*) + (1-\alpha) \cdot \text{load}(d) \big).
    \]
    \STATE Connect all members of $C$ to $d^*$, and connect $d^*$ directly to $g^*$ in FIFO order.
\ENDFOR
\end{algorithmic}
\end{algorithm}

\textbf{Tradeoffs.} Clustered composition aggregates traffic and alleviates gateway congestion by introducing an intermediate layer of aggregation. However, the cluster heads themselves may become bottlenecks if their assigned load grows too large. The choice of $k$ directly influences this balance: too few clusters increase the burden on individual heads, while too many clusters dilute the benefits of aggregation and may reintroduce gateway congestion.

\subsubsection{Parallel Composition}
Parallel composition establishes multiple disjoint relay chains from entry drones to gateways. We assume that each drone forwards to only one successor and does not split its traffic. Under this assumption, parallelism arises from the swarm as a whole: different groups of drones form separate chains operating concurrently. A long sequential relay chain is therefore a special case of parallel composition, where only one path is active.

\begin{algorithm}[h!]
\caption{Parallel Composition Assignment}
\label{alg:parallel}
\begin{algorithmic}[1]

\STATE Compute required number of paths: 
\[
k = \max\!\left( |G|,\; \left\lceil \frac{|U| \lambda}{\mu} \right\rceil \right),
\]
where $|U|$ is the total number of devices, $\lambda$ is the per-device arrival rate, $\mu$ is the per-drone service rate.

\STATE Initialize an empty set of paths $P^*$ and mark all drones as unassigned.

\vspace{0.2cm}
\item[] \textbf{(Gateway-anchored initialization)}
\FOR{each gateway $g \in G$}
    \STATE Select entry drone
    \[
    d^* = \arg\min_{d \in D_\text{entry}} 
    \alpha \cdot \text{dist}(d, g) 
    + (1-\alpha) \cdot \text{load}(d),
    \]
    over unassigned drones.
    \STATE Start a new path $p = [d^*, g]$, add it to $P^*$.
    \STATE Mark $d^*$ as assigned.
\ENDFOR

\vspace{0.2cm}
\item[] \textbf{(Additional path creation if $k > |G|$)}
\IF{$k > |G|$}
    \STATE Build list $\mathcal{L}$ of all pairs $(d, g)$ where $d$ is unassigned:
    \[
    \text{cost}(d,g)=\alpha \cdot \text{dist}(d,g) + (1-\alpha) \cdot \text{load}(d).
    \]
    \STATE Sort $\mathcal{L}$ by cost ascending.
    \FOR{the top $(k - |G|)$ pairs in $\mathcal{L}$}
        \STATE Create new path $p = [d, g]$ and add it to $P^*$.
        \STATE Mark $d$ as assigned.
    \ENDFOR
\ENDIF

\vspace{0.2cm}
\item[] \textbf{(Assign remaining drones to paths)}
\WHILE{there exists an unassigned drone}
    \STATE For each unassigned drone $d$ and each path $p \in P^*$ compute:
    \[
    \text{cost}(d,p)=
    \alpha \cdot \text{dist}(d,p)
    + (1-\alpha) \cdot \text{load}(p),
    \]
    where $\text{load}(p)$ is updated dynamically as members join.
    \STATE Select the pair $(d^*, p^*)$ with minimum cost.
    \STATE Append $d^*$ to the end of path $p^*$.
    \STATE Update $\text{load}(p^*)$.
    \STATE Mark $d^*$ as assigned.
\ENDWHILE

\end{algorithmic}
\end{algorithm}

Algorithm~\ref{alg:parallel} proceeds in four stages. First, it computes the required number of parallel paths from the total demand, respecting drone capacities, while enforcing a lower bound equal to the number of gateways [Lines 1–2]. Second, each gateway initializes one path by attaching the entry drone that minimizes a weighted cost combining proximity and current load, identical to the cost structure used in the clustered strategy. These gateway-anchored choices form the initial set of paths [Lines 3–7]. If the required number of paths exceeds the number of gateways, the algorithm selects additional starting points. It evaluates every unassigned drone against every gateway, computes the same weighted cost, sorts all drone–gateway pairs, and selects the best pairs until the target number of paths is reached [Lines 8–15]. Finally, all remaining drones are assigned to the existing paths. Assignment is performed iteratively and load-aware: at each step, the algorithm connects the drone–path pair that yields the lowest updated cost, where the path's load increases dynamically as drones join it. This ensures that later decisions account for the actual, evolving utilization of each path [Lines 16–22].

\textbf{Tradeoffs.} Parallel composition enhances performance by distributing traffic across multiple relay chains, reducing congestion on any single path. 
The number of parallel paths plays a central role: too few paths concentrate load and increase end-to-end delay, while too many paths fragment traffic and raise coordination and control complexity. 

\subsection{Queuing-Theory Based Heuristic for Composition Selection}
\label{sec:queuing}


The core challenge in SNaaS composition is selecting drone orchestrations that meet latency and stability guarantees under dynamic traffic and deployment conditions. To address this, we use an $\mathrm{M/G/1}$ priority queueing framework, which captures congestion-dominated multi-hop aerial forwarding, supports heterogeneous drones and traffic classes, and enables end-to-end delay evaluation by aggregating per-hop delays \cite{gross2008fundamentals}. The framework maps real-time swarm telemetry to queueing parameters to analytically assess candidate compositions.

\paragraph{Modeling assumptions}
Each drone $d_j$ (entry, relay, or gateway) is modeled as an $\mathrm{M/G/1}$ queue with two traffic classes:
\begin{itemize}
    \item \textbf{High-priority (control) packets}: arrival rate $\lambda_j^{(c)}$, service-time distribution with mean $\bar{X}_j^{(c)}$ and second moment $\mathbb{E}[(X_j^{(c)})^2]$.
    \item \textbf{Low-priority (data) packets}: arrival rate $\lambda_j^{(d)}$, service-time distribution with mean $\bar{X}_j^{(d)}$ and second moment $\mathbb{E}[(X_j^{(d)})^2]$.
\end{itemize}

Let $\lambda_j = \lambda_j^{(c)} + \lambda_j^{(d)}$ be the total arrival rate.
The total load at drone $d_j$ is
\[
\rho_j = \lambda_j^{(c)}\bar{X}_j^{(c)} + \lambda_j^{(d)}\bar{X}_j^{(d)}, 
\qquad \text{stability requires } \rho_j < 1.
\]

The server uses \emph{non-preemptive priority}, with control packets always served ahead of data packets.

\paragraph{Per-node delay}
Define for node $d_j$ the residual service time
\[
R_j = \frac{\lambda_j^{(c)}\,\mathbb{E}[(X_j^{(c)})^2] + 
        \lambda_j^{(d)}\,\mathbb{E}[(X_j^{(d)})^2]}
        {2 (1 - \rho_j)}.
\]

\textbf{High-priority (control) delay.}
Control packets wait only for residual work in service:
\[
W^{(c)}_j = \frac{R_j}{1 - \rho_j^{(c)}},
\qquad
D^{(c)}_j = W^{(c)}_j + \bar{X}^{(c)}_j,
\]
where 
\[
\rho_j^{(c)} = \lambda_j^{(c)} \bar{X}_j^{(c)}.
\]

\textbf{Low-priority (data) delay.}
Data packets wait for: (i) residual time, (ii) all queued control packets, and (iii) control arriving during the waiting period:
\[
W^{(d)}_j = \frac{R_j}{(1 - \rho_j^{(c)})(1 - \rho_j)},
\qquad
D^{(d)}_j = W^{(d)}_j + \bar{X}^{(d)}_j.
\]

\paragraph{Mapping strategies to queueing networks}
Each composition induces a routing pattern and therefore a set of arrival rates $\{\lambda_j^{(c)}, \lambda_j^{(d)}\}$ at each drone:
\begin{itemize}
    \item \emph{Direct:} multiple entry drones send control and data directly to gateways (parallel $\mathrm{M/G/1}$ servers).
    \item \emph{Clustered:} entries $\rightarrow$ cluster heads $\rightarrow$ gateways, creating a two-layer priority queueing network.
    \item \emph{Parallel:} several disjoint relay chains toward gateways; each path is a series of $\mathrm{M/G/1}$ priority queues.
\end{itemize}

\paragraph{End-to-end latency}
Let $\pi$ be a path from an entry drone to a gateway:
\[
\pi = (d_s, d_1, \dots, d_k, g).
\]
The end-to-end latency for traffic class $x \in \{c,d\}$ is
\[
L^{(x)}(\pi) = \sum_{\ell \in \pi} D^{(x)}_\ell.
\]

Given a composition $c$ with path set $\Pi_c$ and traffic fractions $\omega_\pi$:
\[
L_{\text{avg}}(c)= \sum_{\pi\in \Pi_c} \omega_\pi L^{(d)}(\pi), 
\qquad
L_{\max}(c)= \max_{\pi\in \Pi_c} L^{(d)}(\pi).
\]
Mission-critical SLAs typically evaluate worst-case latency; consumer-grade tasks may use average latency.

\paragraph{Decision rule}
Given measured $\lambda_j^{(c)},\lambda_j^{(d)}$ and service distributions, select the composition
\[
\begin{aligned}
c^{*} &= \arg\min_{c\in\mathcal{C}} L_{\text{SLA}}(c) \\
\text{s.t.}\quad 
& \rho_j < 1,\;\forall j,\\
& |U_j| \le C_d\ \text{(coverage/capacity constraint)}.
\end{aligned}
\]

Algorithm~\ref{alg:qheur} describes the main steps in the selection process. [Line~1] collects the current swarm telemetry (per-node control and data arrival rates and service-time statistics) and enumerates the set of candidate compositions $\mathcal{C}$ (e.g., direct, clustered, and parallel variants instantiated for the current topology). For each candidate composition $c \in \mathcal{C}$ [Lines~2--7], [Line~3] derives the resulting routing pattern by mapping device traffic to end-to-end paths $\Pi_c$ and computing the induced arrival rates at every drone along those paths. Given these per-node rates, Line~4 evaluates the priority $\mathrm{M/G/1}$ queueing model at each node, computing the residual service time $R_j$ and the class-specific delays $D^{(c)}_j$ and $D^{(d)}_j$. [Line~5] then aggregates these per-node delays into an end-to-end latency metric $L_{\text{avg}}(c)$ or $L_{\max}(c)$, depending on whether the SLA is defined in terms of average or worst-case latency. In [Line~6], any composition whose utilization violates the stability condition $\rho_j < 1$ at any node is marked infeasible and excluded from further consideration. Finally, [Line~8] selects the composition $c^{*}$ that minimizes the SLA-relevant latency metric $L_{\text{SLA}}(c)$ among all feasible candidates, and returns it as the chosen orchestration plan.

\begin{algorithm}[h]
\caption{Priority $\mathrm{M/G/1}$ Guided Strategy Selection}
\label{alg:qheur}
\begin{algorithmic}[1]
\STATE Collect telemetry: arrival rates $\lambda_j^{(c)},\lambda_j^{(d)}$, service statistics, and enumerate $\mathcal{C}$.
\FOR{each composition $c\in\mathcal{C}$}
    \STATE Derive routing $\Rightarrow$ $\Pi_c$ and per-node arrival rates.
    \STATE Compute $R_j$, $D^{(c)}_j$, $D^{(d)}_j$ for all nodes.
    \STATE Evaluate $L_{\text{avg}}(c)$ or $L_{\max}(c)$ depending on SLA.
    \STATE Mark $c$ infeasible if $\rho_j\ge 1$ for any $j$.
\ENDFOR
\STATE Return $c^{*}=\arg\min_{c\in\mathcal{C}} L_{\text{SLA}}(c)$.
\end{algorithmic}
\end{algorithm}

\subsection{SLA Compliance and Stability Enforcement}

If queueing analysis shows that no candidate composition satisfies queue stability ($\rho_j < 1$) or meets the SLA latency bound, the controller activates a stability enforcement (SE) stage, since both conditions are mandatory contractual requirements. The SE module iteratively applies corrective actions, recomputing utilizations and end-to-end latency after each step, and halts as soon as all queues satisfy $\rho_j \le \rho_{\max} < 1$ and latency compliance is restored, thereby minimizing reconfiguration overhead while ensuring stable and SLA-compliant service delivery.

\paragraph*{Procedure}

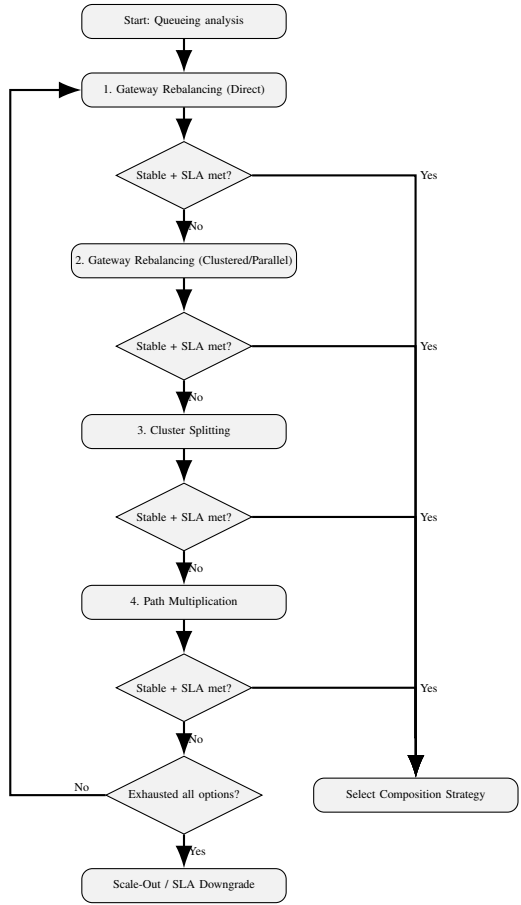
\begin{figure}[h]
\centering
\begin{tikzpicture}[node distance=1cm, scale=0.45, every node/.style={transform shape, font=\normalsize}]

\node (start) [block] {Start: Queueing analysis};
\node (direct) [block, below=of start] {1. Gateway Rebalancing (Direct)};
\node (dec1) [decision, below=of direct] {Stable + SLA met?};
\node (cluster) [block, below=of dec1] {2. Gateway Rebalancing (Clustered/Parallel)};
\node (dec2) [decision, below=of cluster] {Stable + SLA met?};
\node (split) [block, below=of dec2] {3. Cluster Splitting};
\node (dec3) [decision, below=of split] {Stable + SLA met?};
\node (path) [block, below=of dec3] {4. Path Multiplication};
\node (dec4) [decision, below=of path] {Stable + SLA met?};

\node (exhaust) [decision, below=of dec4] {Exhausted all options?};
\node (scale)   [block, below=of exhaust] {Scale-Out / SLA Downgrade};
\node (select)  [block, right=1.5cm of exhaust] {Select Composition Strategy};

\draw [arrow] (start) -- (direct);
\draw [arrow] (direct) -- (dec1);
\draw [arrow] (dec1) -- node[right]{No} (cluster);
\draw [arrow] (dec1.east) -| node[right]{Yes} ++(4.8,0) |- ([yshift=1.2cm]select.north) -- (select.north);

\draw [arrow] (cluster) -- (dec2);
\draw [arrow] (dec2) -- node[right]{No} (split);
\draw [arrow] (dec2.east) -| node[right]{Yes} ++(4.8,0) |- ([yshift=1.2cm]select.north) -- (select.north);

\draw [arrow] (split) -- (dec3);
\draw [arrow] (dec3) -- node[right]{No} (path);
\draw [arrow] (dec3.east) -| node[right]{Yes} ++(4.8,0) |- ([yshift=1.2cm]select.north) -- (select.north);

\draw [arrow] (path) -- (dec4);

\draw [arrow] (dec4) -- node[right]{No} (exhaust);
\draw [arrow] (dec4.east) -| node[right]{Yes} ++(4.8,0) |- ([yshift=1.2cm]select.north) -- (select.north);

\draw [arrow] (exhaust) -- node[right]{Yes} (scale);
\draw [arrow] (exhaust.west) -- ++(-2.8,0) node[above,pos=0.25]{No} |- (direct.west);

\end{tikzpicture}
\caption{Flowchart of the SLA Compliance and Stability Enforcement procedure.}
\label{fig:stability_flow}
\end{figure}

The enforcement logic, summarized in Fig.~\ref{fig:stability_flow}, proceeds through four ordered edits, where each edit is applied once in sequence; if SLA compliance is not achieved, the procedure then iteratively cycles through the same edits until all corrective options are exhausted.

1) \textbf{Gateway rebalancing (direct).}
The controller first applies direct composition adjustments, as they are minimally disruptive and affect only routing, not the swarm structure. Entry drones are reassigned to alternate gateways by adapting the weighting factor $\alpha_g$ to discourage heavily loaded gateways and favor lighter ones. This allows connections to farther gateways when needed. If the resulting mapping satisfies gateway stability but still violates the SLA, or remains unstable, the controller escalates to the next enforcement step.

2) \textbf{Gateway rebalancing (clustered/parallel).}  
This step redistributes aggregate traffic by reassigning cluster heads or relay chains to alternate gateways when direct reassignment is insufficient. The controller adapts the gateway weighting factor $\alpha_g$ to penalize saturated gateways and favor underloaded ones while preserving internal composition. If the resulting mapping satisfies gateway stability and SLA latency constraints, enforcement stops; otherwise, it proceeds to the next step.

3) \textbf{Cluster splitting.}  
If gateway rebalancing fails, the controller splits overloaded clusters by dividing any cluster with $\lambda_h^{\text{agg}} > \mu_h$ into two subclusters to reduce per-head utilization. The split increases the number of clusters ($k \rightarrow k+1$) using the same k-means spatial and load-aware criteria as in composition, followed by local head reselection. If recomputed stability or latency constraints are still violated, the controller proceeds to the next step.

4) \textbf{Path multiplication.}
Cluster splitting is applied before path multiplication, as it addresses overload at the aggregation level, while path multiplication restructures relay paths. If congestion persists, the controller splits highly utilized relay chains and recomposes them into parallel paths, dividing traffic to reduce per-relay load and improve stability. The updated paths are then evaluated for queue stability and end-to-end latency, with further enforcement applied if violations remain.

\textbf{Scale-out or SLA downgrade.}  
If neither stability nor the SLA latency target can be achieved after all preceding corrections, the controller initiates a scale-out action or triggers SLA renegotiation. 
In a scale-out action, new drones or gateways are provisioned to expand service capacity and restore compliance. If resource addition is infeasible, the controller requests an SLA downgrade, i.e., loosening the latency bound or reducing the traffic demand, to reestablish a feasible operating region under the current infrastructure.

\section{Experiments and Results}
This section evaluates the proposed framework implemented on Google Colab (Python) using a high-RAM CPU runtime ($\approx$50 GB RAM) on a Google Compute Engine backend.

\subsection{Dataset}

Experiments employ the AERPAW UAV-based Signal Dataset \cite{dickerson2025aerpaw}, which provides real-world air-to-ground (A2G) received-signal strength (RSS) measurements at a 3.3 GHz carrier frequency for three UAV altitudes (40 m, 70 m, 100 m). The dataset was collected at five fixed ground nodes, supplying spatial diversity in the measured link quality. For each altitude, RSS samples collected across all ground nodes and time were aggregated to compute a single mean RSS value. During simulation, channel bandwidth was treated as a control parameter and swept between 5–20 MHz. For each altitude–bandwidth configuration, signal-to-noise ratios (SNRs) were computed using a bandwidth-dependent thermal noise model, and per-UAV service rates ($\mu$) were deterministically estimated using a Shannon-capacity-based throughput proxy scaled by a PHY efficiency factor ($\eta = 0.6$). These $\mu$ values were used as fixed service parameters in the $\mathrm{M/G/1}$ priority queues.

\subsection{Experimental Parameters}

Unless stated otherwise, simulations assume a transmit power of $P_t = 20$~dBm, a data packet size of 1~kB, a control packet size of 32~B, a carrier frequency of 3.3~GHz, and a duration of 120~s. Thermal noise power \cite{goldsmith2005wireless} is computed as:
\[
N = -174 + 10\log_{10}(B) + NF,
\]
where $B$ denotes the channel bandwidth and $NF$ is the receiver noise figure \cite{goldsmith2005wireless}. We fix the UAV altitude at 40~m and the channel bandwidth at 5~MHz.


\subsection{Request Generation}

A total of 100 synthetic consumer requests are generated per evaluation bin, each specifying an SLA with average latency $L_{\text{SLA}}$ and stability threshold $\rho < 1$. Each request also defines the number of user devices to be served and their allocation to preassigned entry drones, respecting each drone’s capacity $C_d$. User devices are uniformly randomly distributed within a $100 \text{m} \times 100 \text{m}$ service area, while UAV positions are also randomized. To preserve controlled comparisons across operating points, results are grouped into bins along the x-axis (e.g., SLA targets or number of devices), and the 100 requests within each bin are varied by a small perturbation of $\pm10\%$ around the bin’s nominal value. This captures realistic variations in user density and load distribution across the swarm under consistent spatial constraints. The SNaaS controller dynamically composes UAV services to satisfy these requirements. Per-user traffic arrivals are modeled as a Poisson process, reflecting independent user activity and aligning with standard assumptions in $\mathrm{M/G/1}$ service models.



\subsection{Scaling and Topology}
Real AERPAW link distributions are used to parameterize the simulations by mapping link quality to per-link service rates. We assume interference is mitigated through standard wireless resource management techniques (e.g., TDMA/FDMA-based orthogonal channel allocation, power control, frequency reuse, and directional antennas), and detailed interference-aware optimization in SNaaS is left for future work. We evaluate three scales representing disaster-response scenarios:

\begin{itemize}
    \item \textbf{Small:} 10 entry UAVs, 3 gateway
    \item \textbf{Medium:} 25 entry UAVs, 5 gateways
    \item \textbf{Large:} 100 entry UAVs, 10 gateways
\end{itemize}



\subsection{Experimental Evaluation}

\begin{figure}[h!]
    \centering
    \includegraphics[width=\linewidth]{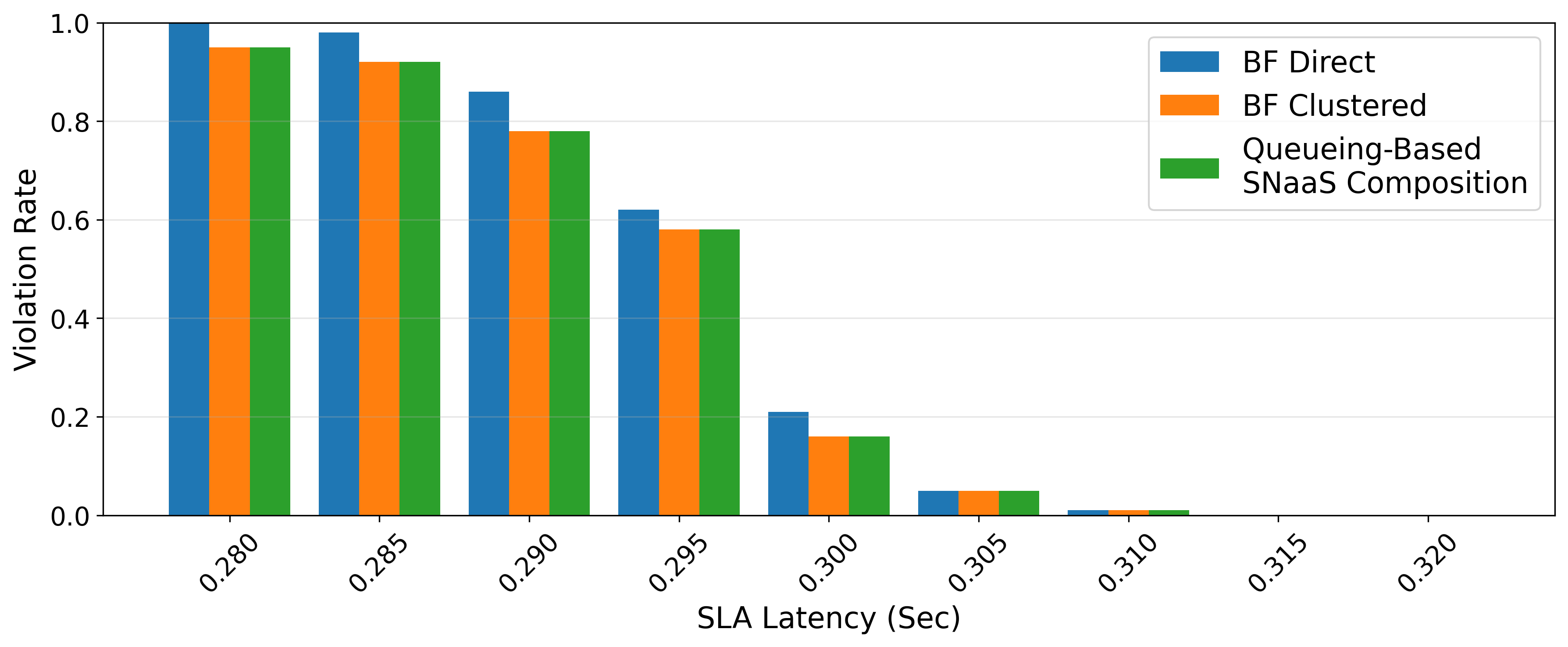}
    \caption{Violation rate versus requested SLA latency for a small-scale swarm.}
    \label{fig:viol_rate_vs_sla}
\end{figure}

\begin{figure}[h!]
    \centering
    \includegraphics[width=0.9\linewidth]{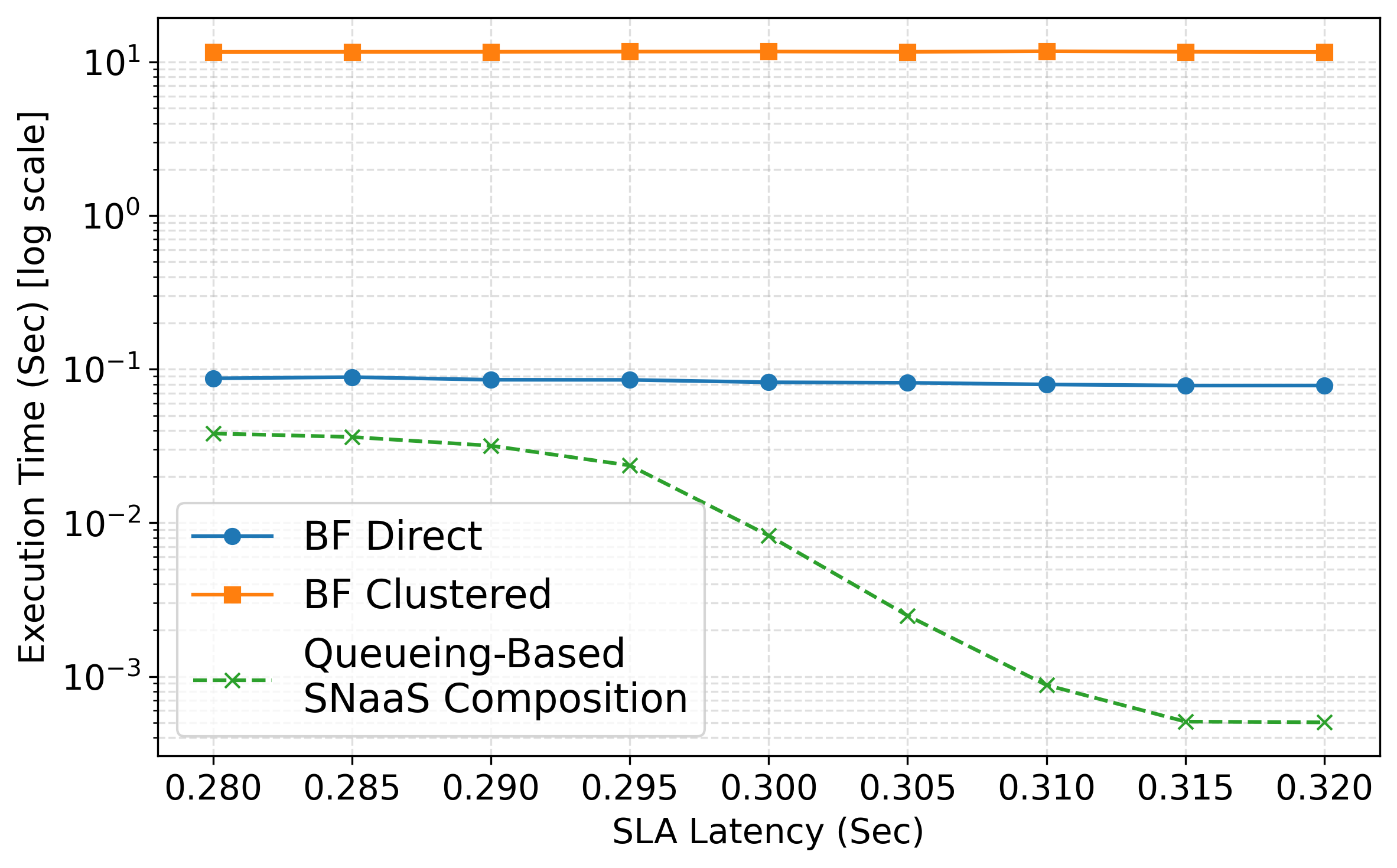}
    \caption{Average execution time per request versus SLA latency (log scale).}
    \label{fig:time_vs_sla}
\end{figure}

\paragraph{\textbf{Experiment 1: Violation Rate vs. SLA Latency}}
We study the impact of SLA strictness on system compliance by measuring the violation rate as a function of the requested latency bound (Fig. \ref{fig:viol_rate_vs_sla}). For each requested SLA latency, 100 requests are generated on a small-scale swarm. We compare our proposed queueing-theory-based enforcement against two brute-force baselines: brute-force Direct and brute-force Clustered compositions, each enumerating all feasible combinations up to a cap of 200 candidates; a brute-force Parallel baseline is omitted due to its prohibitive computational cost, and parallel composition is not generally useful for small-scale swarms, as further demonstrated in subsequent experiments. The results show that relaxing the SLA (higher allowable latency) consistently reduces violations for all methods. Brute-force Direct exhibits the highest violation rate, as directly connected devices overload gateways and create queueing bottlenecks. In contrast, brute-force Clustered and our approach achieve nearly identical and significantly lower violation rates, indicating that clustered composition is the dominant structure at this scale and that our method successfully selects the same solutions identified by brute force. Importantly, as shown in the accompanying runtime figure (Fig. \ref{fig:time_vs_sla}), our approach reaches this performance with substantially lower computation time, highlighting its efficiency advantage.

\begin{figure}[h!]
    \centering
    \includegraphics[width=0.8\linewidth]{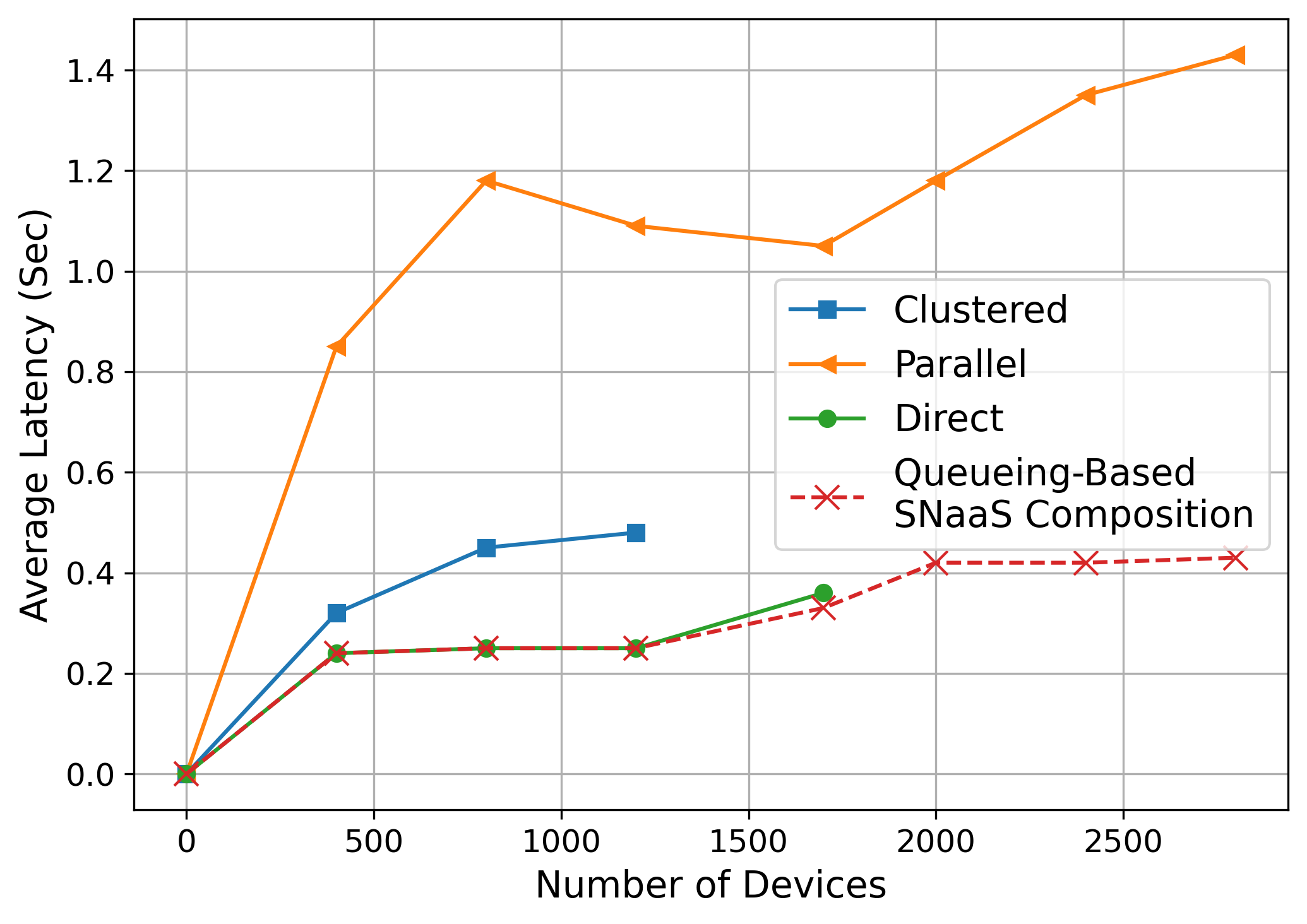}
    \caption{Average latency versus number of devices for a medium-scale swarm.}
    \label{fig:latency_vs_devices}
\end{figure}

\paragraph{\textbf{Experiment 2: Latency vs. Number of Devices}}
This experiment highlights the importance of the stability enforcement step by reporting the average latency as the number of connected devices increases in a medium-scale swarm (Fig.~\ref{fig:latency_vs_devices}). We fix the requested SLA latency and progressively increase the offered load by adding devices, averaging results over multiple Monte Carlo runs. We compare direct, clustered, and parallel compositions without enforcement against the proposed framework with enforcement enabled. Minor local fluctuations in the curves may appear due to the controlled randomness in request generation as explained earlier, but the overall trend remains the key observation. The results show that parallel composition without enforcement achieves the highest latency; although additional hops distribute load and maintain stability as the number of devices increases, this comes at the cost of significantly higher end-to-end delay. In contrast, direct and clustered compositions without enforcement become unstable early as device count grows, due to excessive load on gateways or cluster heads; clustered fails earlier than direct because the number of clusters are not optimized as the weight for the cost function is not optimized ($\alpha =0.5$). Our proposed framework consistently outperforms all baselines by dynamically switching between composition strategies as load increases while enforcing stability constraints, and by fine-tuning the cost function weight ($\alpha$) to achieve the lowest average latency. This adaptive behavior and strategy selection are further analyzed in the following experiment.\looseness=-1

\paragraph{\textbf{Experiment 3: Strategy Selection Behavior Under Increasing Load}}

\begin{figure}[h!]
    \centering
    \includegraphics[width=\linewidth]{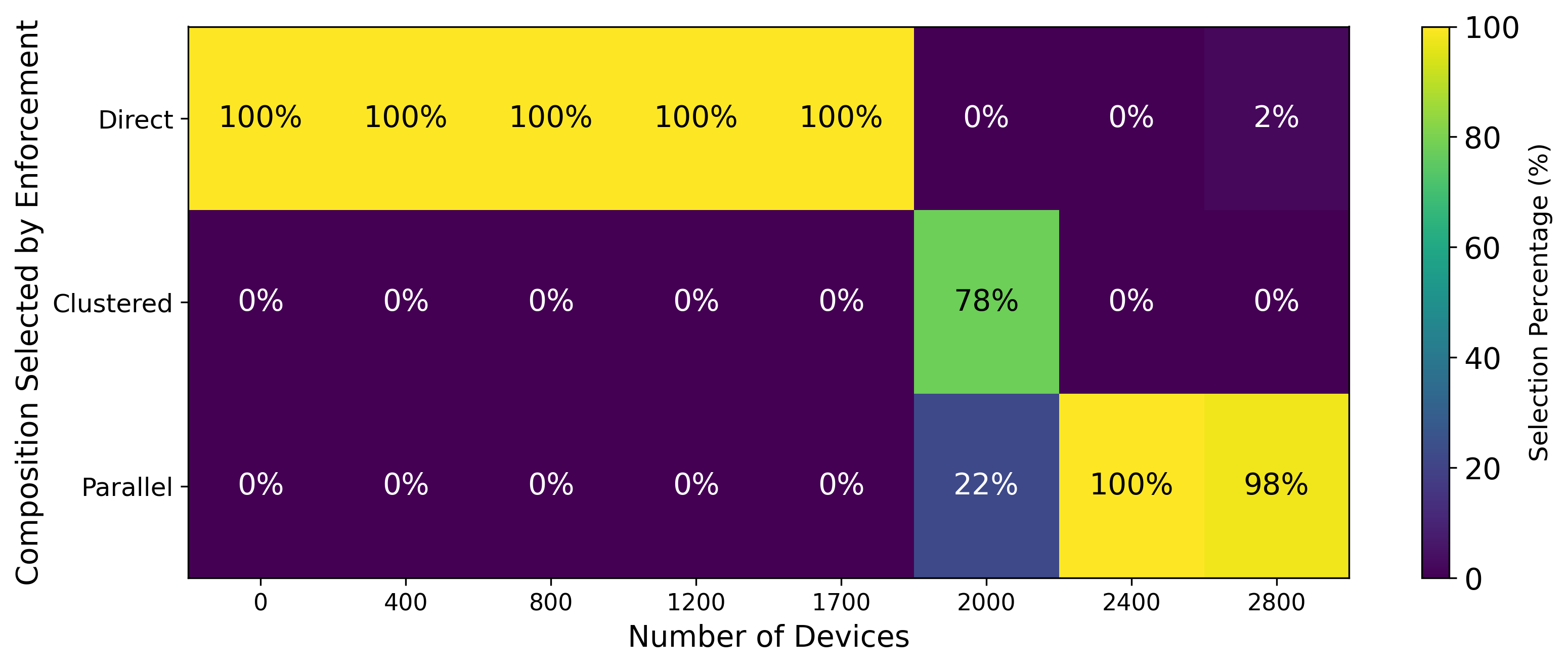}
    \caption{Heatmap of composition strategy selection frequency versus number of devices.}
    \label{fig:selection_heatmap}
\end{figure}

This experiment analyzes how the SLA-aware enforcement mechanism selects composition strategies as device load increases (Fig.~\ref{fig:selection_heatmap}). At low load, direct composition dominates due to minimal latency. As load grows and stability degrades, the framework shifts to clustered composition, and at high load increasingly selects parallel composition to distribute traffic. This progression demonstrates adaptive strategy selection that balances latency and stability to satisfy SLA constraints. The small residual direct selection at high load is attributable to randomness.

\begin{figure}[h!]
    \centering
    \includegraphics[width=0.8\linewidth]{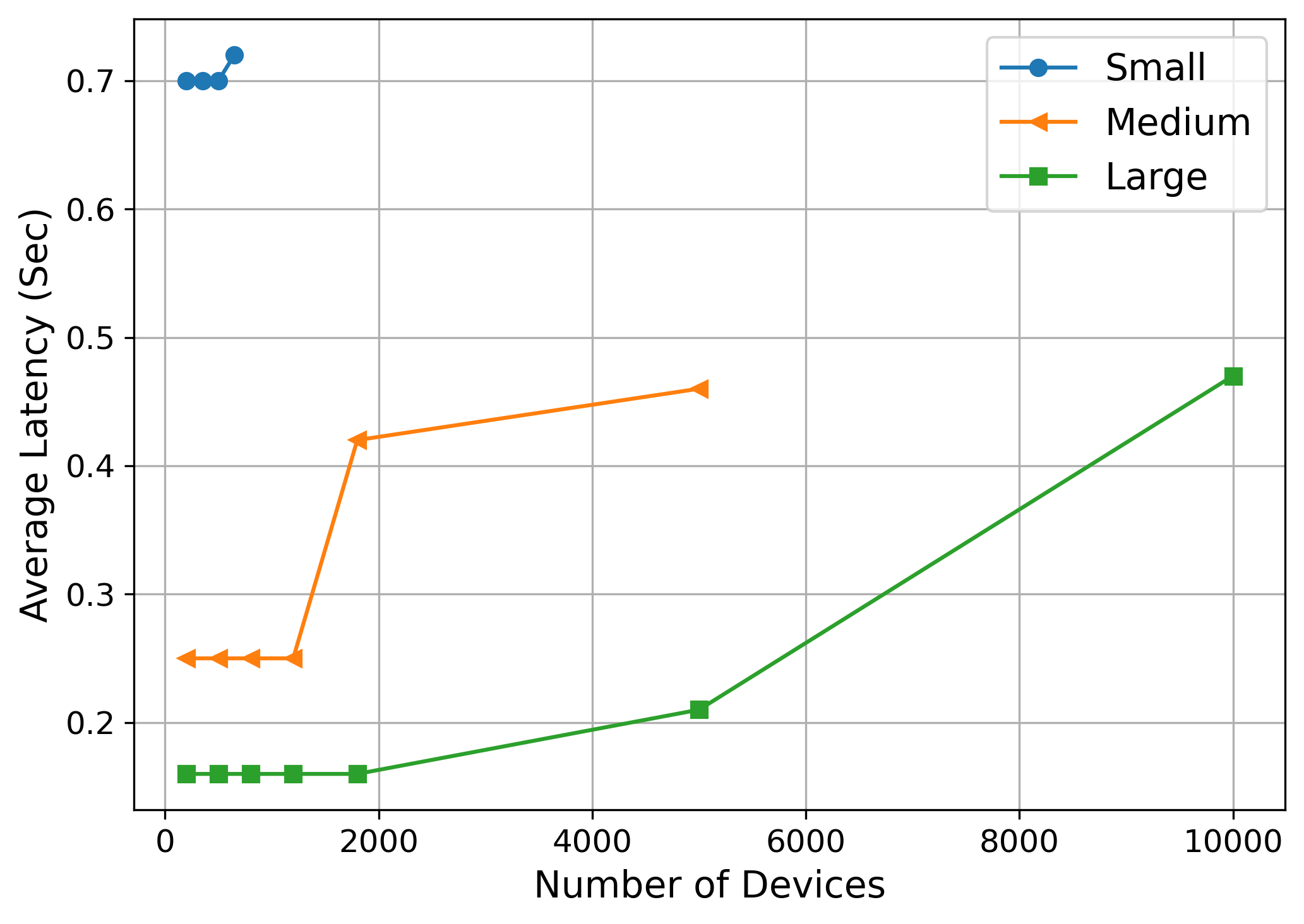}
    \caption{Average latency versus number of devices under SLA enforcement for different swarm scales (small/medium/large).}
    \label{fig:scale_vs_devices}
\end{figure}

\paragraph{\textbf{Experiment 4: Impact of Swarm Size on Latency Under Increasing Load}}
This experiment evaluates how swarm size affects scalability and latency as the number of connected devices increases under a fixed SLA latency requirement (Fig.~\ref{fig:scale_vs_devices}). Unlike the earlier definition of small, medium, and large scenarios, where swarm scale and device counts were jointly varied, here the same device load is applied across all swarm scales to isolate the effect of swarm size alone. Using the proposed enforcement mechanism, we measure the average queueing latency while progressively increasing device counts for three swarm scales (small, medium, large), averaging over Monte Carlo trials with randomized device placements. The results show that larger swarms sustain stable operation for substantially higher device counts and achieve lower latency, since additional drones provide more service capacity and enable better load distribution across gateways and relays. This trend highlights the performance–cost tradeoff in swarm deployment: increasing the number of drones improves feasibility and reduces latency under higher device demand, but at the expense of greater deployment cost, enabling operators to determine the minimum swarm size required to satisfy a given SLA.\looseness=-1

\section{Conclusion and Future Work}
This paper introduces Swarm Network-as-a-Service (SNaaS) as a service-oriented abstraction for on-demand, SLA-driven connectivity using UAV swarms. It models drone-to-device and drone-to-drone interactions as composable services, formalizes atomic and composite SNaaS services, and presents an SOA–SDN-inspired architecture that addresses service composition under latency and stability constraints.

The paper proposes direct, clustered, and parallel composition strategies together with a queueing-theory-based selection and enforcement framework that analytically evaluates stability and end-to-end latency using a priority $\mathrm{M/G/1}$ model. Experimental results based on real AERPAW data show near brute-force performance at significantly lower computational cost, smooth adaptation across load regimes, and clear performance–cost tradeoffs as swarm size scales. Future direction can extend SNaaS with swarm-driven coordination, where drones adapt routes using local queue and latency feedback to enable self-organization, faster response, and resilient, SLA-compliant operation. In this setting, each drone pursues local optimization goals that collectively converge toward globally efficient and SLA-compliant network states.

\section*{Acknowledgments}
The work is partly funded by the King Abdullah University of Science and Technology (KAUST) Global Fellowship Program - Award No. RFS-KGFP2024-6784.



\bibliographystyle{IEEEtran} 
\bibliography{references}

\vfill

\end{document}